\documentclass[oneside, a4paper, onecolumn, 11pt]{article}
\usepackage[left=2cm,top=2cm,bottom=2cm,right=2cm]{geometry}
\usepackage{graphicx} 
\usepackage{epstopdf, epsfig} 
\usepackage{amsmath} 

\usepackage{color}
    \definecolor{IMV1}{rgb}{0.64, 0.0, 0.0}
\usepackage[colorlinks,linkcolor={IMV1},citecolor={IMV1},urlcolor={IMV1}]{hyperref} 

\usepackage{siunitx}
\usepackage{bm} 
\usepackage{derivative} 
\usepackage[noabbrev]{cleveref} 

\usepackage{natbib}
\setcitestyle{authoryear,open={(},close={)}}

\newcommand{\X}{\ensuremath{\mathbf{x}}}
\newcommand{\Vel}{\ensuremath{\mathbf{u}}}
\newcommand{\Vort}{\ensuremath{\boldsymbol{\omega}}}
\newcommand{\Str}{\ensuremath{\boldsymbol{\Gamma}}}
\newcommand{\Rad}{\ensuremath{\boldsymbol{r}}}

\begin{document}
\begin{center}

\bigskip
{\Large \textbf{Toward Meshless Turbulent Flow Simulation: LES-Integrated Vortex Particle Method}} \\
\bigskip

Flavio A. C. Martins$^{1,*}$, Alexander van Zuijlen$^{2}$ and Carlos J. Sim\~{a}o Ferreira$^{1}$

$^{1}$ Faculty of Aerospace Engineering, Flow Physics and Technology Department, Wind Energy ($^2$Aerodynamics) Section, Delft University of Technology \\ Kluyverweg 1, Delft, The Netherlands

$^*$ F.M.Martins@tudelft.nl

\bigskip
{\Large \textbf{ABSTRACT}} \\
\bigskip
\end{center}

Recent developments in vortex particle methods for simulating three-dimensional incompressible flows are presented. A lightweight, dynamic Large-Eddy Simulation model is tested, featuring a dynamic procedure that relies solely on Lagrangian information and requires minimal auxiliary computation to update the model constant. The method employs a high-order algebraic kernel which enables direct, analytical expressions for conservation laws, the strain-rate tensor, and quadratic velocity diagnostics. Viscous diffusion is modeled using the core-spreading technique. The particle method is assessed with respect to kinematics and the conservation of energy, helicity, and enstrophy in vortex ring and leapfrogging vortex ring scenarios, both unperturbed and perturbed. The results indicate that the kinematics and flow diagnostics are accurately captured using relatively sparse particle distributions, effectively resolving the dynamics of the unstable flow phase. However, after the onset of instabilities, the sub-grid-scale methodology becomes strongly dependent on particle regularization to stabilize the flow solution.




\section{INTRODUCTION} \label{section:introduction}


Vortex methods offer an alternative paradigm for computational fluid dynamics (CFD), formulating the incompressible Navier-Stokes equations in velocity-vorticity form and solving them within a fundamentally Lagrangian framework. Among these, the Vortex Particle Method (VPM) discretizes the vorticity field using particles, each carrying a smooth radial basis function to construct a continuous vorticity field~\citep{leonard1980vortex}. This mesh-free approach enables high-fidelity transport of vortical structures over long distances while circumventing the Courant-Friedrichs-Lewy (CFL) condition and minimizing artificial dissipation~\citep{mimeau2021review}. As a result, VPM has demonstrated significant advantages in terms of computational efficiency and accuracy, often outperforming mesh-based methods by orders of magnitude for equivalent fidelity~\citep{yokota2013petascale}.

Despite these strengths, the development of robust three-dimensional Large-Eddy Simulation (LES) models tailored to VPMs has remained limited. While foundational contributions to Lagrangian turbulence modeling date back to the 1970s~\citep{chorin1973numerical, christiansen1973numerical, milinazzo1977calculation}, more recent efforts have largely focused on Direct Numerical Simulation (DNS)~\citep{yokota2011vortex} or conditionally stable LES formulations~\citep{barba2005advances, gharakhani2005lagrangian}. As a result, fully three-dimensional LES models that retain the core strengths of VPM are still scarce.

Current LES strategies in VPM can be broadly categorized into two classes: (i) fully Lagrangian approaches, which maintain a completely mesh-free formulation~\citep{gharakhani2003application}, and (ii) semi-Lagrangian approaches, which project particle data onto background grids to compute quantities such as vorticity gradients, vortex stretching, or viscous diffusion~\citep{mimeau2021comparison}. While the latter benefit from established Eulerian techniques, they reintroduce many of the limitations--such as mesh generation, numerical dissipation, and CFL constraints--that vortex methods are designed to avoid. Fully Lagrangian methods, on the other hand, retain these advantages but often face challenges in defining dynamic filtering procedures, handling anisotropy, and computing spatial derivatives without auxiliary meshes.

Motivated by these opportunities, this work proposes a fully Lagrangian LES formulation for VPM that operates independently of background meshes. The approach extends the dynamic Smagorinsky model~\citep{deardorff1970numerical, lilly1992proposed}, reformulated within a Lagrangian framework for defining and estimating filter scales, and adapted to a high-order algebraic kernel~\cite{winckelmans1993contributions}. Viscous diffusion is modeled using a core-spreading technique~\citep{greengard1985core}, which inherently preserves the mesh-free and Lagrangian character of the method. The resulting LES-enhanced vortex particle method is validated on canonical test cases involving isolated and leapfrogging vortex rings, with emphasis on conservation properties and the accurate reproduction of vortical structure kinematics.

The remainder of this paper is organized as follows. Section~\ref{section:methodology} introduces the LES-VPM, detailing its key numerical components and the formulation of a scale-consistent LES filter within the particle-based framework. Section~\ref{section:results} presents a two-tier validation strategy based on canonical vortex ring problems: an isolated vortex ring, used to assess energy dissipation, core spreading, and stability under axisymmetric evolution; and a pair of interacting (leapfrogging) vortex rings, which introduces non-axisymmetric instabilities, mutual induction, and turbulence-induced breakdown, thereby increasing both the modeling and computational challenges.

\section{Foundations of the LES Vortex Particle Method} \label{section:methodology}


\subsection{Governing Equations of Incompressible Flow}
\label{section:governing_eq}

The present vortex particle method solves the filtered vorticity transport equation for incompressible turbulent flows,\footnote{For clarity, conventional filtering accents (e.g., tildes or hats) that denote resolved LES quantities are omitted.} given by:

\begin{align}
    \nabla^2 \Vel &= - \nabla \times \Vort, \label{equation:Poisson} \\
    \frac{D\Vort}{Dt} &= \Vort \cdot \nabla \Vel + \nu \nabla^2 \Vort + \mathbf{g},
\label{equation:vorticity_transport}
\end{align}

where $\Vel$ is the velocity field, $\Vort$ is the vorticity field, $\nu$ is the kinematic viscosity, and $\mathbf{g}$ denotes the subfilter-scale (SGS) vorticity torque accounting for unresolved interactions between velocity and vorticity.

To close the system defined by \eqref{equation:Poisson} and \eqref{equation:vorticity_transport}, a vorticity-based eddy-viscosity model is adopted \citep{cottet2003vorticity, mansfield1999dynamic}:

\begin{equation}
    \mathbf{g} = \nabla \cdot (\nu_T \nabla \Vort),
\label{equation:unresolved_momentum}
\end{equation}

where $\nu_T$ is the turbulent eddy viscosity, modeled as

\begin{equation}
    \nu_T = C_\omega^2 \Delta^2 |\mathbf{S}|,
\label{equation:turbulent_nu}
\end{equation}

where \(C_\omega\) is a dynamically computed model constant, \(\Delta\) is the LES filter width and $\mathbf{S}$ is the velocity strain-rate tensor. As will be discussed, the adopted high-order algebraic kernel allows for exact and computationally inexpensive evaluations of the velocity field gradients, without requiring auxiliary grids. With the SGS model defined by \eqref{equation:unresolved_momentum} and \eqref{equation:turbulent_nu}, the vorticity transport equation \eqref{equation:vorticity_transport} can be rewritten as:

\begin{equation}
    \frac{D\Vort}{Dt} = \Vort \cdot \nabla \Vel + (\nu + \nu_T) \nabla^2 \Vort.
\label{equation:vorticity_transport2}
\end{equation}

\subsection{Vortex Particle Solver Formulation}
\label{section:vpm_formulation}

The vorticity field is discretized using regularized vortex particles. Each particle $p$ carries a vector strength $\Str^p$, and the continuous vorticity field is approximated as:

\begin{equation}
    \Vort(\X,t) \approx \sum_p \Str^p(t) \, \zeta_\sigma\left(\X - \X^p(t)\right),
\label{equation:vorticity_approximated}
\end{equation}

where $\X$ is an arbitrary point in the flow field, $\X^p$ is the position vector of the $p$-th particle, and $\zeta_\sigma(|\X - \X^p|) \equiv \zeta(|\X - \X^p|/\sigma)$ is the mollifier, i.e., a smooth kernel function with characteristic core size $\sigma$. In the present VPM model, we adopt the high-order algebraic kernel introduced by \citet{winckelmans1993contributions}:

\begin{equation}
    \zeta(\rho) = \frac{15}{8\pi} \, \frac{1}{(\rho^2 + 1)^{7/2}},
\label{equation:kernel}
\end{equation}

with $\rho \equiv |\X - \X^p| / \sigma$. This kernel provides closed-form expressions for the velocity field and quadratic flow diagnostics (e.g., kinetic energy, helicity, and enstrophy), offering a convenient analytical structure for LES modeling, while retaining convergence properties similar to those of a Gaussian kernel.

A regularized particle representation of the continuous velocity field is obtained by solving the Poisson equation \eqref{equation:Poisson}, which, via the Green’s function formulation, leads to the aerodynamic-theory analogue of the Biot–Savart law:

\begin{equation}
    \Vel_\sigma(\X,t) = \sum_p \mathbf{K}_\sigma(\X - \X^p(t)) \times \Str^p(t) + \mathbf{u}_{\infty},
\end{equation}

where $\Vel_\sigma$ is the regularized velocity field, and the regularized Biot-Savart kernel is given by $\mathbf{K}_\sigma(\Rad) = -q_\sigma(\Rad) \, \Rad / |\Rad|^3$, with $\Rad = \X - \X^p$ and

\begin{equation}
     q(\rho) = \frac{1}{4 \pi} \frac{\rho^3 \left( \rho^2 + \tfrac{5}{2} \right)}{(\rho^2 + 1)^{5/2}}.
\end{equation}

In the current VPM formulation, Eq.~\eqref{equation:vorticity_transport} is solved using a fractional-step method: the vortex stretching and tilting (first term on the right-hand side) and the viscous diffusion (second term) are evaluated separately from the material derivative. The evolution of the vortex particles is thus described by:

\begin{align}
    \frac{d\X^p(t)}{dt} &= \Vel_\sigma \left( \X^p(t), t \right) + \mathbf{u}_{\infty}, \label{equation:update_positions} \\
    \frac{d\Str^p(t)}{dt} &= \left( \Str^p(t) \cdot \nabla^T \right) \Vel_\sigma \left( \X^p(t), t \right), \label{equation:update_strengths_adv} \\
    \frac{d\Str^p(t)}{dt} &= \nabla \cdot \left( (\nu + \nu_T^p) \nabla \Str^p \right), \label{equation:update_strengths_diff}
\end{align}

where Eq.~\eqref{equation:update_strengths_diff} corresponds to the transposed form of Fickian diffusion applied to the vortex strength.

For viscous diffusion, the core spreading method (CSM) is adopted (see \citep{leonard1980vortex} and \citep{cottet2000vortex}). In this method, the diffusive operator in Eq.~\eqref{equation:update_strengths_diff} is replaced by a time-dependent broadening of the regularization kernel $\zeta_\sigma$, such that the equation is not solved explicitly. Instead, the kernel's support size $\sigma$ evolves in time as:

\begin{equation}
    \Str^p(t) \text{ is held constant during diffusion, while } (\sigma^p)^2(t) = (\sigma^p)^2(t-\Delta t) + C_\nu (\nu + \nu_T^p)\Delta t,
\label{eq:core_spreading_sigma}
\end{equation}

where $\Delta t$ is the time step size and $C_\nu = 256/45$ is the viscous diffusion constant determined by the authors based on how the second moment of the vorticity distribution evolves under viscous diffusion. The relation in Eq.~\ref{eq:core_spreading_sigma} is derived from the analytical solution of the diffusion equation for a Gaussian distribution and captures viscous effects through kernel broadening.

Using the high-order algebraic kernel from Eq.~\eqref{equation:kernel}, the following closed-form expressions are obtained for the $p$-th particle's motion and strength update, depending on the remaining $q$ particles:

\begin{align}
    \frac{d \X^p(t)}{dt} &= -\frac{1}{4\pi} \sum_q \left[ A (\X^p - \X^q) \times \Str^q \right], \label{equation:update_positions2} \\
    \frac{d \Str^p(t)}{dt} &= \frac{1}{4\pi} \sum_q \left[ A (\Str^p - \Str^q) + 3B \left( \Str^p \cdot ((\X^p - \X^q) \times \Str^q) \right) (\X^p - \X^q) \right], \label{equation:update_strengths2}
    \\
    (\sigma^p)^2(t) &= (\sigma^p)^2(t-\Delta t) + C_\nu(\nu + \nu_T^p)\Delta t.
\end{align}

Here, the coefficients $A$ and $B$ are derived analytically as functions of $|\X^p - \X^q|$ and $\sigma$:

\begin{align}
    A &= \frac{ | \X^p - \X^q |^2 + \tfrac{5}{2}\sigma^2 }{ \left( | \X^p - \X^q |^2 + \sigma^2 \right)^{5/2} }, \\
    B &= \frac{1}{ \left( | \X^p - \X^q |^2 + \sigma^2 \right)^{7/2} }
\end{align}

\subsection{Linking LES and Vortex Particle Filters}
\label{section:linking_LES_VPM}

The present filtering approach is based on associating the resolution limit of the particle discretization with the particle core radius $\sigma$, and treating Eqs.~\eqref{equation:Poisson} and~\eqref{equation:vorticity_transport} as ``particle-filtered'' equations. The notion of filter size is therefore standardized by relating $\sigma$ to the width of an equivalent box filter, denoted $\Delta$. To this end, we equate the energy content of a spherical top-hat filter of width $\Delta$ to that of the vortex particle regularization kernel $\zeta$. The top-hat kernel is defined as:

\begin{equation}
    b(r) =
    \begin{cases}
    \displaystyle \frac{6}{\pi \Delta^3}, & r \leq \frac{\Delta}{2}, \\
    0, & \text{otherwise},
    \end{cases}
\end{equation}

and its associated filter energy is given by:

\begin{equation}
    E_b = \int_0^{\Delta/2} |b(r)|^2 \, 4\pi r^2 \, dr.
\end{equation}

Meanwhile, the energy content of the radially symmetric mollifier $\zeta$, used to regularize the vortex particle representation, is expressed as:

\begin{equation}
    E_\zeta(c) = \int_0^{\infty} |\zeta(r; c)|^2 \, 4\pi r^2 \, dr,
\label{equation:energies}
\end{equation}

where $c \equiv \Delta/\sigma$ is a scaling factor determined by the condition that the particle-filter energy content matches that of the spherical top-hat filter, i.e., $E_\zeta(c) = E_b$. This energy-matching framework establishes a consistent connection between conventional LES modeling parameters and the regularization scale employed in the vortex particle method. For the high-order algebraic kernel described in Eq.~\eqref{equation:kernel}, the energy-equivalent scale factor is approximately $c \approx 2.366$. 

Equation~\eqref{equation:turbulent_nu} can thus be rewritten as:

\begin{equation}
    \nu_T^p = (C_\omega^p \Delta^p)^2 |\mathbf{S}^p| = (C_\omega^p \sigma^p c)^2 |\mathbf{S}^p|,
\end{equation}

where $|\mathbf{S}^p|$ denotes the regularized particle strain-rate tensor. The tensor norm is defined as $\left| \mathbf{\bullet} \right| = \sqrt{2 \, \mathbf{\bullet} : \mathbf{\bullet}}$, with $:$ denoting the Frobenius inner product (double contraction). This is determined by computing the spatial gradients of the adopted high-order kernel function, and $C_\omega^p$ is the particle-based filter constant.

\subsection{Dynamic Eddy Viscosity Filtering}
\label{section:selective_filtering}

The dynamic filter constant $C_\omega^2$ for each particle $p$ is computed using the Germano identity \citep{lilly1992proposed}, adapted to vortex particle methods:

\begin{equation}
C_\omega^2 = \frac{\langle \mathbf{L} : \mathbf{M} \rangle}{\langle \mathbf{M} : \mathbf{M} \rangle + \epsilon},
\end{equation}

where $\mathbf{L}$ is the resolved sub-grid stress tensor, and $\mathbf{M}$ is the model stress tensor at the particle location. The angle brackets $\langle \bullet : \bullet \rangle$ denote a local spatial average over neighboring particles within the filter radius $\Delta$, and $\epsilon$ is a small regularization constant that prevents division by zero. The tensors $\mathbf{L}$ and $\mathbf{M}$ are defined as
$ \mathbf{L} = \overline{\overline{\mathbf{M}}}_{2\Delta} - \overline{\mathbf{M}}_{\Delta} $ with:

\begin{align}
&\mathbf{M} = -2 (\Delta^p)^2 \left| \mathbf{S}^p \right| \mathbf{S}^p, \\
&\overline{\mathbf{M}}_{\Delta} = \frac{1}{N_{\Delta}} \sum_{k \in \mathcal{N}_{\Delta}} \mathbf{M}_k, \\
&\overline{\overline{\mathbf{M}}}_{2\Delta} = -2 (2\Delta^p)^2 \left| \overline{\mathbf{S}}_{2\Delta} \right| \overline{\mathbf{S}}_{2\Delta}.
\end{align}

Here, $\Delta^p$ and $2\Delta^p$ are the filter widths at the particle location. $\mathbf{S}^p$ is the local strain-rate tensor at particle $p$, and $\overline{\mathbf{S}}_{2\Delta}$ is the average strain-rate tensor over neighboring particles within radius $2\Delta^p$. $N_{\Delta}$ is the number of neighbors within radius $\Delta^p$ (excluding the particle itself), and $\mathcal{N}_{\Delta}$ denotes the set of these neighbors. Under these definitions, $\mathbf{L}$ captures the resolved inter-scale stress between filter widths $\Delta^p$ and $2\Delta^p$, while $\mathbf{M}$ provides a local sub-filter stress model. Neighboring particles are efficiently identified via a fast tree-code algorithm \citep{maneewongvatana1999analysis,contributors2020scipy}.

\section{Results}\label{section:results}
\subsection{Vortex Ring Test Case}

\begin{figure}[htbp]
    \centering
    \includegraphics[width=11cm]{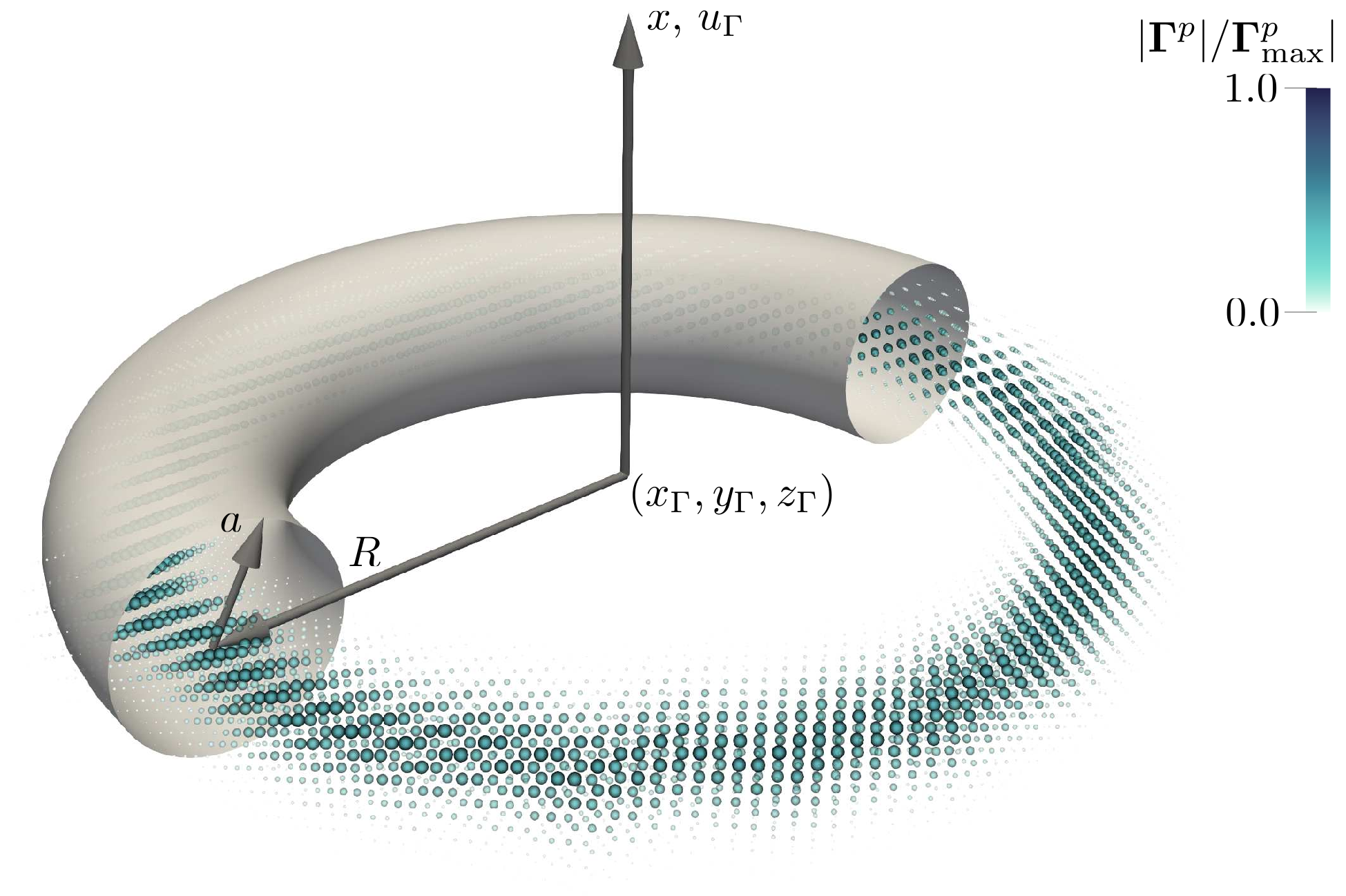}
    \caption{Vortex ring with azimuthal perturbations represented by vortex particles, colored by normalized strength magnitude. The self-induced velocity direction, $u_\Gamma$, is highlighted. An unperturbed ring, depicted as a gray surface, is overlaid for reference.}
    \label{fig:Vortex_Ring_Flow}
\end{figure}

The first test case involves a vortex ring flow, a classical validation scenario for both vorticity advection and viscous diffusion. Vortex rings are primarily characterized by their self-propelling speed, $u_\Gamma$, which decreases over time as vorticity diffuses due to viscosity. The case considered here features a vortex ring with a strength-based Reynolds number of $Re_\Gamma = \Gamma_0 / \nu = 7500$, where $\Gamma_0$ is the circulation of the ring (not to be confused with the particle strengths $\Str^p$), and an initial core thickness of $a_0/R_0 = 0.2$. Here, $R_0$ is the ring's initial radius, and the subscript $0$ identifies properties at $t=0$. The core thickness is defined by $a_0 = 4 \nu (t_0 - t_{-\nu})$. A schematic of the model is shown in Fig.~\ref{fig:Vortex_Ring_Flow}.

The initial vorticity distribution in the ring's core follows a Gaussian profile:

\begin{equation}
    \Omega(\Rad,t) = \frac{\Gamma_0}{4 \pi \nu (t - t_{-\nu})} 
    \exp\left( -\frac{\|\Rad\|^2}{4 \nu (t - t_{-\nu})} \right),
\end{equation}

where $\Rad$ denotes the position in a local coordinate system centered at the ring.

Two scenarios are simulated: (i) an unperturbed vortex ring, and (ii) a perturbed ring with azimuthal perturbations in its radius introduced via:

\begin{equation}
   R_0 \leftarrow R_0 \sum_{n_W=1}^{24} (1 + \delta_R \cos(n_W \theta)), 
\end{equation}

where $\delta_R/a_0 = 0.025$ and $\theta$ is the azimuthal coordinate. The Widnall modes used here are based on the instability modes described by \citet{widnall1974instability}. The perturbation amplitude was chosen to be the smallest that triggered instabilities in the ring evolution for the adopted average inter-particle distance, $h$.

The VPM representation of the vorticity field introduces numerical diffusion due to the particles' finite core size, $\sigma^p$. To compensate, an “anti-diffusion” time shift $t_{-\nu}$ is used to ensure that the analytical diffusion profile matches the particle approximation \citep{barba2004vortex}. For the chosen kernel, this is given by: $t_{-\nu} = \sigma_0^{p,2} / (C_{\nu} \nu)$.

A second-order Runge-Kutta scheme is employed to integrate all time-dependent quantities, including particle positions and strength updates. Particles are initialized on a regular hexagonal grid with uniform spacing $h$, and are assigned strengths using a radial basis fit to the target vorticity distribution, $\Omega(\X^p, 0) \mathcal{V}^p$, where $\mathcal{V}^p$ is the volume of each cell in the grid. Particles with local vorticity below $5\%$ of the global maximum are discarded to reduce computational cost. The initial particle core sizes are computed using $\sigma_0^p = 4\sqrt{h}/5$ \citep{barba2004vortex}. The time step is set as $\Delta t = 20 h^2/\Gamma_0$, which ensures that the kinetic energy variation between steps remains below $0.1\%$, considered a good temporal resolution.

A resolution independence test was performed by comparing solutions at different values of $h/a_0$. Results were deemed resolution-independent for $h/a_0 < 0.3$, based on agreement with analytical diffusion profiles and convergence across successive particle resolutions. A value of $h/a_0 = 0.25$ was selected for the simulations, corresponding to approximately 9,000 vortex particles At this particle resolution, each time step required approximately 3.9 seconds on average when executed on an NVIDIA RTX 3060 Laptop GPU. The LES filter width is defined as $\Delta = h$, i.e., the minimum resolvable scale corresponds to the average inter-particle distance. The associated test filter, by definition, encompasses all particles within a distance of $2\Delta$ from the $p$-th particle.

\begin{figure}[htbp]
\centering
\includegraphics[width=12.5cm]{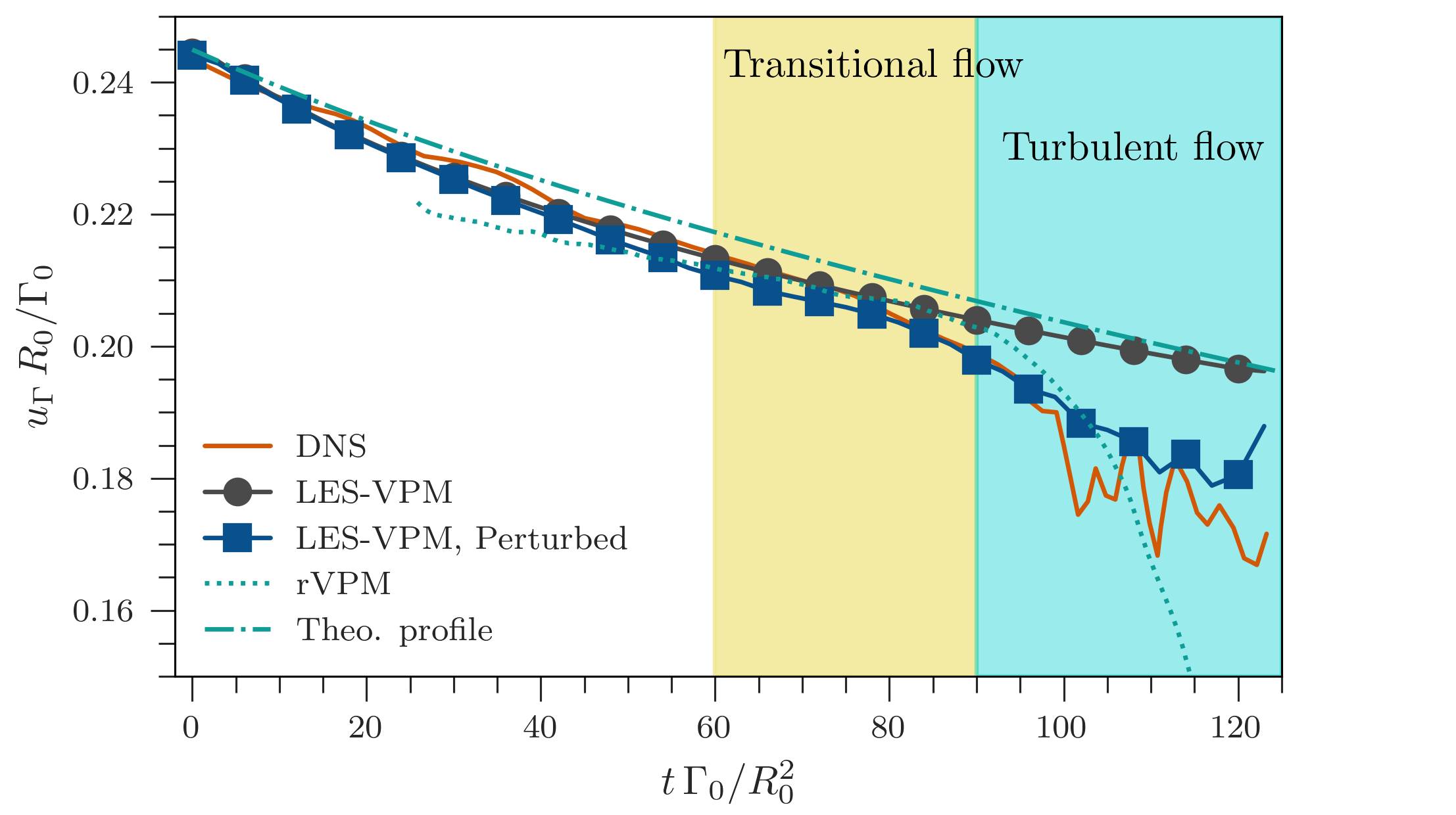}
\caption{Translational velocity of the vortex ring from the present LES-VPM model, compared with DNS results, theoretical predictions, and the rVPM method. The white, yellow, and blue regions represent the laminar, transitional, and turbulent flow regimes, respectively.}
\label{fig:Vortex_Ring_Speed}
\end{figure}

Figure~\ref{fig:Vortex_Ring_Speed} shows the normalized translational velocity of the vortex ring over time. The present LES-VPM method is compared against a semi-empirical theoretical prediction from \citet{archer2008direct}, their DNS results (using a spectral code with an azimuthal perturbation of $\delta_R = 2 \times 10^{-4}$ computed on a mesh with approximately 100 million cells), and the rVPM method of \citet{alvarez2022reviving} (with 500,000 vortex particles). The velocity history of the unperturbed case closely follows the theoretical prediction throughout the entire simulated period, maintaining a difference of $\sim2\%$ or less relative to the latter.

For the perturbed case, the ring's translational velocity exhibits three distinct behaviors corresponding to different flow phases. These are identified in Fig.~\ref{fig:Vortex_Ring_Speed} as laminar (white region), transitional (yellow), and turbulent (blue) flows~\citep{archer2008direct}. During the laminar phase ($t \Gamma_0 / R_0 < 60$), the ring speed evolves closely in line with both the theoretical and DNS profiles. In the transitional period ($60 < t \Gamma_0 / R_0 < 90$), increased turbulent diffusion leads to an earlier and more pronounced deceleration, with the perturbed LES-VPM closely matching the results from rVPM and DNS. Finally, in the fully turbulent phase ($t \Gamma_0 / R_0 > 90$), the velocity decreases significantly, and noticeable fluctuations in ring speed emerge, consistent with turbulence-induced instabilities.

\begin{figure}[htbp]
    \centering
    \includegraphics[width=15cm]{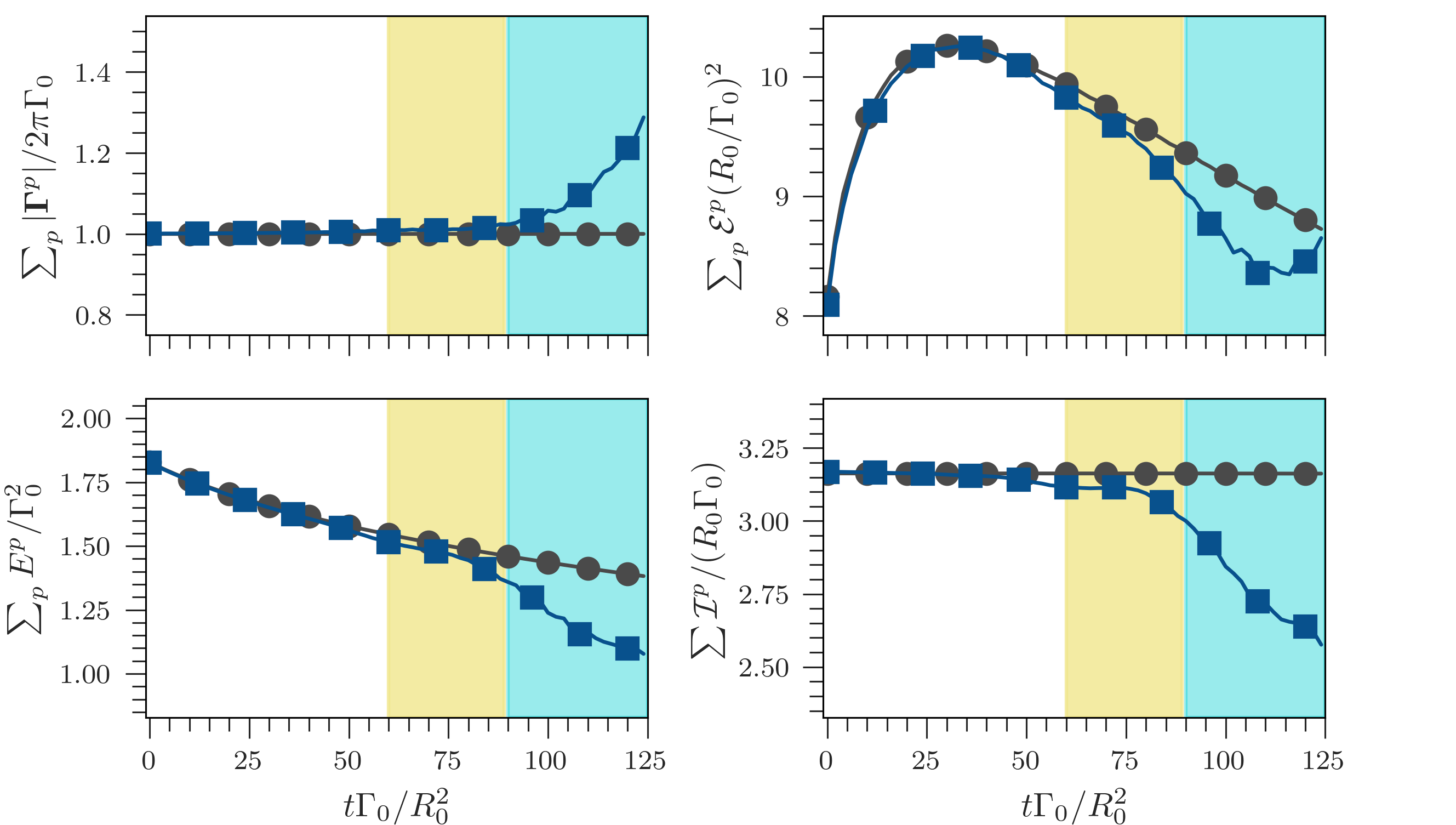}
    \caption{Evolution of vortex ring diagnostics for perturbed and unperturbed cases: (top row) circulation and enstrophy; (bottom row) kinetic energy and linear impulse. The white, yellow, and blue regions represent the laminar, transitional, and turbulent flow regimes, respectively.}
    \label{fig:Vortex_Ring_Diagnostics}
\end{figure}

Figure~\ref{fig:Vortex_Ring_Diagnostics} presents the evolution of integral diagnostics--total circulation ($\Str^p$), enstrophy ($\mathcal{E}^p$), kinetic energy ($E^p$), and linear impulse ($\mathcal{I}^p$)--for both perturbed and unperturbed rings. The results are overlaid with the three flow regimes--laminar, transitional, and turbulent--as defined by the authors, shown in white, yellow, and blue, respectively. Initially, the vortex core undergoes a transient equilibration phase, due to the finite-width Gaussian vorticity profile used as the ring's initial condition being only an approximate solution to the Navier-Stokes equations. This manifests as a temporary increase in enstrophy for $t \Gamma_0 / R_0 < 30$, consistent with the observations of \citet{archer2008direct}, who referred to this early adjustment as a relaxation phase of the initial solution. After equilibration, the enstrophy decreases as vorticity is gradually shed from the ring.

Kinetic energy decays at a rate, $\Delta E / \Delta t$, consistent with the viscous dissipation term $-\nu \mathcal{E}^p$, remaining within 30\% of the latter. This is a reasonable approximation for the assessed LES-VPM methodology in an unbounded viscous flow, where the energy balance simplifies to $\partial E / \partial t + \nu \mathcal{E} \approx 0$. In the unperturbed case, both total circulation and linear impulse are nearly conserved (within 0.5\%), indicating that the LES scheme and core-spreading mechanism preserve the divergence-free condition effectively. However, for the perturbed ring, divergence errors grow after the onset of strong instabilities ($t \Gamma_0 / R_0 > 100$) due to the lack of regularization and increasingly non-uniform particle distribution, which degrade the accuracy of the vorticity representation.

\subsection{Leapfrogging Vortex Rings}
\label{section:leapgrogging}

\begin{figure}[htbp]
    \centering
    \includegraphics[width=11cm]{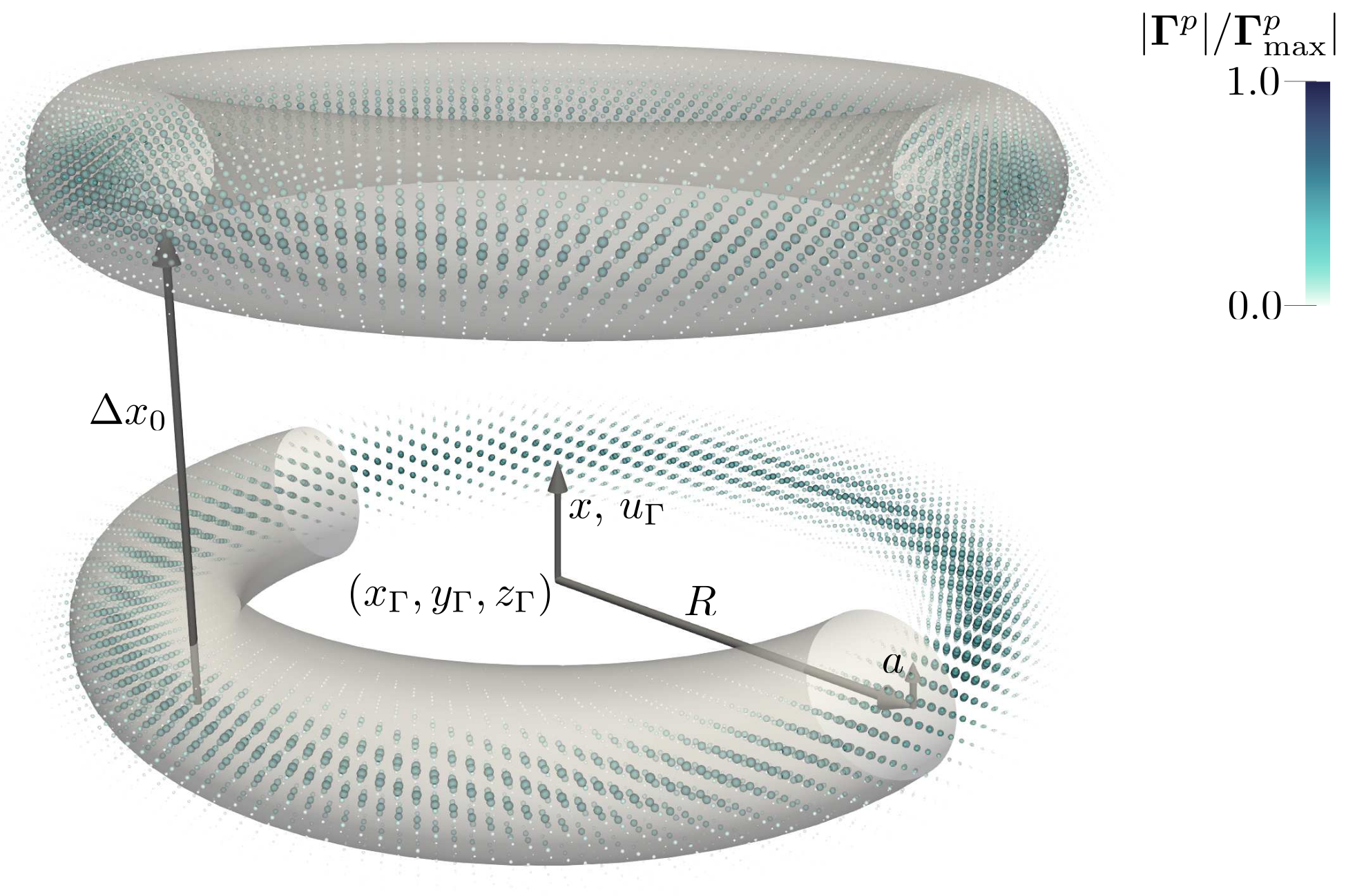}
    \caption{Two leapfrogging vortex rings represented by strength-colored vortex particles. The rings are radially perturbed to induce the onset of flow instabilities during the turbulent flow regime. The non-disturbed shape of the vortex rings is overlaid as gray surfaces for reference.}
    \label{fig:Leapfrogging_Rings_Flow}
\end{figure}

The interaction of two vortex rings traveling in the same direction, often observed as a repeated ``leapfrogging'' motion, serves as a validation case for models incorporating vortex stretching and complex vortex dynamics. During this phenomenon, the rings exhibit cyclical stretching and contraction as they successively pass through one another. For this study, we simulate two vortex rings at a strength-based Reynolds number of $Re_\Gamma = 3000$, each with an initial core thickness of $a_0/R_0=0.1$, and initially separated by a distance of $x/R_0=2$. The configuration of the system is depicted in Fig. \ref{fig:Leapfrogging_Rings_Flow}.

The system is initialized with a particle field having a uniform distribution spacing $h$ of $h/a_0=0.2$, which results in approximately 17,000 vortex particles. At this particle resolution, each time step required approximately 12 seconds on average. This higher particle concentration, compared to the single ring case, was necessary to ensure that the anti-diffused initial time, $t-t_{-\nu}$, remains positive. Both rings are similarly perturbed with the first 24 Widnall modes ($n_W \in [1,24]$), each with a perturbation amplitude of $\delta_R/R_0 = 0.025$ and randomly-seeded phases for the perturbation wave components. All other simulation parameters strictly follow those established for the single vortex ring case.

\begin{figure}[htbp]
    \centering
    \includegraphics[width=14cm]{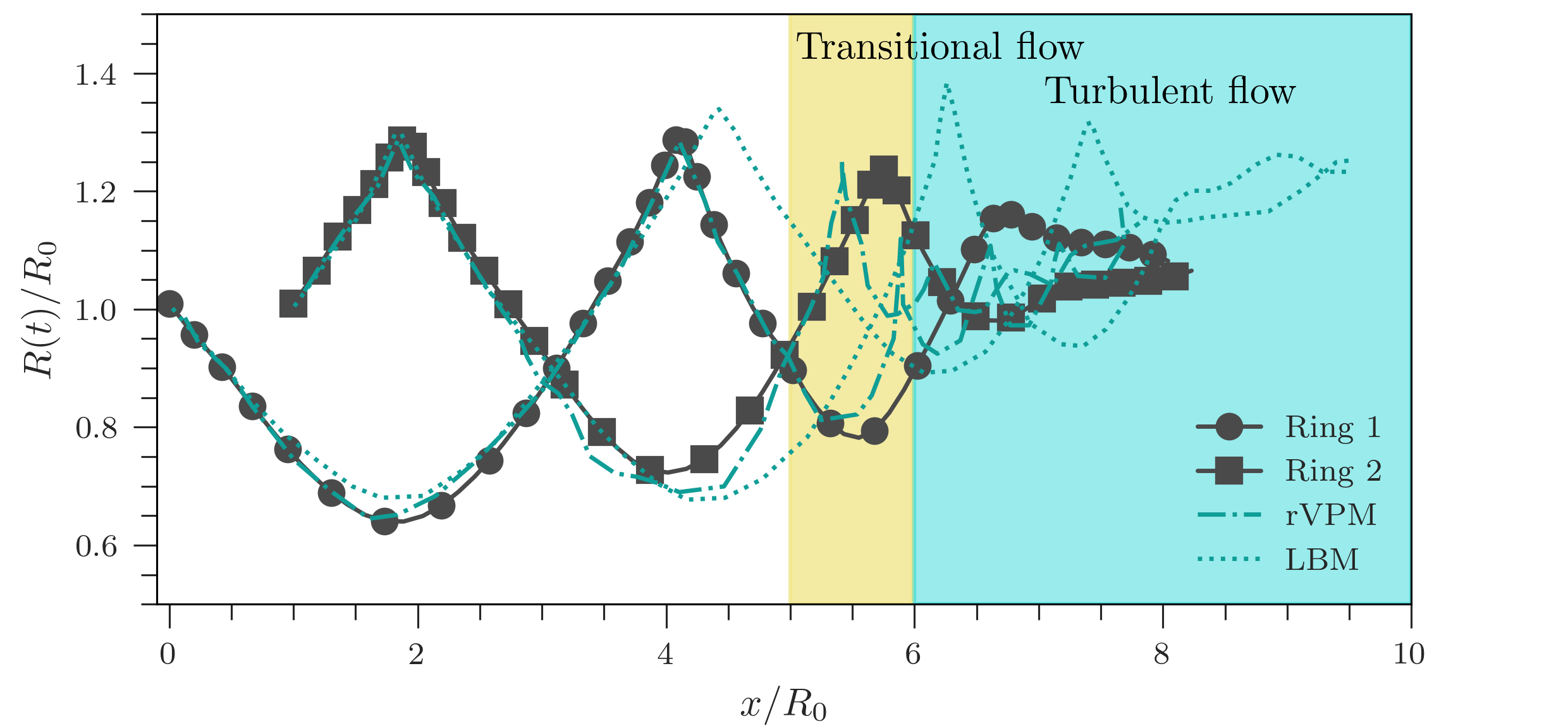}
    \caption{Translational velocity of the leapfrogging vortex rings from the present LES-VPM model, compared with LBM results by \cite{cheng2015leapfrogging} and rVPM results by \cite{alvarez2022reviving}. The white, yellow, and blue regions represent the laminar, transitional, and turbulent flow regimes, respectively.}
    \label{fig:Leapfrogging_Rings_Motion}
\end{figure}

Figure \ref{fig:Leapfrogging_Rings_Motion} illustrates the translational history of the leapfrogging rings in the $(x,R)$ plane. Our results are compared with simulations from the literature: a lattice-Boltzmann method (LBM)-based simulation by \cite{cheng2015leapfrogging}, which used a perturbation of $\delta_R/R_0 = 0.05$ for mode $n_W=8$ and was computed on a system with 15 billion cells, and the results reported by \cite{alvarez2022reviving} for their rVPM model, which employed 600,000 vortex particles. These comparisons clearly demonstrate the well-known contraction-expansion pattern characteristic of leapfrogging ring dynamics.

The authors subdivided the flow regimes into laminar, transitional, and turbulent based on a visual inspection of the leapfrogging patterns in the $R$--$x$ plane, for simplicity. During the laminar regime ($x / R_0 < 5$), flow instabilities have not yet fully developed, and all three methodologies show relatively good agreement in capturing the stretching of the vortical structures--see Fig.~\ref{fig:Leapfrogging_Rings_Motion}. In the subsequent transitional phase, approximately between $5 < x / R_0 < 6$, the increased turbulent viscosity, as modeled by the SGS approach, leads to a deceleration of the system's mean translational velocity and an increase in the frequency of the leapfrogging motion. This diagnostic indirectly assesses whether the current SGS model, on average, produces sufficient turbulent viscosity ($\nu_t$) relative to the LBM code.Finally, in the turbulent flow regime, coherent leapfrogging motion ceases as both vortex rings become highly distorted and eventually merge into a single, disordered vortical structure due to strong nonlinear interactions and enhanced turbulent diffusion.

It is important to note that both the LBM and rVPM methodologies used for comparison do not employ the anti-diffusivity time shift to properly set the initial conditions for the rings. Consequently, they tend to overestimate the initial vorticity distribution by approximately 30\%, despite their finer system resolutions, which leads to an overestimation of the rings' initial speed. Nonetheless, the magnitude of the leapfrogging motion and the onset of instabilities, at $x/R_0>6$, are consistently captured by all three compared methodologies.

Previous tests (not shown here) by the authors indicate that without the initial radial perturbation ($\delta_R$) imposed on the rings, the leapfrogging motion would persist until approximately $8x/R_0$, agreeing more closely with the unperturbed rVPM results from \citet{alvarez2022reviving}. However, the imposed perturbation triggers earlier vorticity breakdown due to the momentum imbalance caused by the randomly-distributed perturbation modes. This behavior aligns more closely with the LBM results regarding the timing of instability onset.

\begin{figure}[htbp]
\centering
\includegraphics[width=16cm]{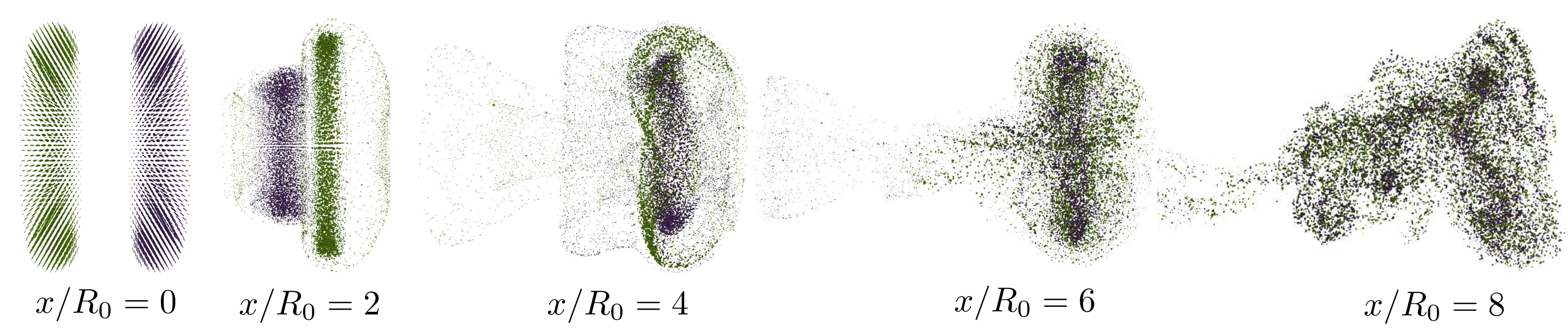}
\caption{Visual evolution of the leapfrogging vortex rings at different normalized axial positions: (a) $x/R_0=0$, (b) $x/R_0=2$, (c) $x/R_0=4$, and (d) $x/R_0=6$. Particles are colored by their respective rings.}
\label{fig:Leapfrogging_Rings_Phases}
\end{figure}

Figure~\ref{fig:Leapfrogging_Rings_Phases} shows the evolution of vortex-ring-tagged particles at normalized axial positions $x/R_0 = 0,\, 2,\, 4,\, 6,\,$ and $8$ (left to right). These visualizations reinforce the trends identified in Fig.~\ref{fig:Leapfrogging_Rings_Motion}. For $x/R_0 < 4$, the dynamics remain largely axisymmetric, with small radial perturbations that do not compromise the overall ring coherence. Between $4 < x/R_0 < 6$, nonlinear amplification of azimuthal and short-wavelength instabilities disrupts the leapfrogging motion, marking the transition to irregular vortex dynamics. Beyond $x/R_0 > 6$, filamentation and reconnection dominate, leading to the breakdown of individual rings. In this turbulent phase--defined heuristically by the authors--vortex cores dissolve, and particles become dispersed and entrained into the surrounding flow, indicating the onset of turbulence-driven mixing.

Although a detailed SGS analysis--e.g., based on velocity gradient tensors, energy spectra, or inter-scale transfer--is beyond the scope of this initial study, the results support the robustness of the proposed LES framework. The model closely matches high-fidelity DNS and rVPM solutions, while remaining numerically stable despite the absence of explicit particle regularization--often considered critical for resolving complex vortex dynamics in VPM. These results suggest that the current LES-VPM approach offers a reliable and computationally efficient tool for simulating unbounded, vortex-dominated turbulent flows.


\section{CONCLUSIONS} \label{section:conclusions}

This study introduced and validated a fully Lagrangian Large-Eddy Simulation framework integrated into a vortex particle method (LES-VPM) for modeling incompressible, vortex-dominated flows. By leveraging a high-order algebraic kernel and a dynamic subgrid-scale model formulated via Lagrangian quantities, the method retains the mesh-free, divergence-free, and transport-accurate features of classical VPM while enhancing its capability to model turbulent dissipation. Validation on two canonical problems--the evolution of a single vortex ring and the leapfrogging of perturbed vortex rings--demonstrated that the LES-VPM approach captures key physical behaviors observed in DNS and advanced vortex methods, including velocity decay, enstrophy evolution, and the onset of instability. These results establish the current LES-VPM framework as a reasonable alternative for high-fidelity simulation of three-dimensional turbulent flows.


\bibliography{references}
\end{document}